\documentclass[12pt]{article}
\usepackage{amssymb}

\makeatletter
\def\numberbysection{\@addtoreset{equation}{section}
 	\def\theequation{\thesection.\arabic{equation}}}
\makeatother

\numberbysection

\newcommand{\be}{\begin{eqnarray}}
\newcommand{\ee}{\end{eqnarray}}
\newcommand{\non}{\nonumber}
\newcommand{\tr}{\mathop{\rm tr}\nolimits}
\newcommand{\id}{\mathbb{I}}

\newcommand{\M}{\mathop{\cal M}\nolimits}

\newcommand{\ch}{\mathop{\rm ch}\nolimits}
\newcommand{\sh}{\mathop{\rm sh}\nolimits}
\newcommand{\tnh}{\mathop{\rm th}\nolimits}
\newcommand{\cth}{\mathop{\rm cth}\nolimits}

\begin{document}

\begin{titlepage}
\strut\hfill
\vspace{.5in}
\begin{center}

\LARGE Generalized $T$-$Q$ relations and the open spin-$s$ XXZ\\
\LARGE chain with nondiagonal boundary terms\\[1.0in]
\large Rashad Baiyasi\footnote{email:ribaiyas@svsu.edu} and Rajan Murgan\footnote{e-mail: rmurgan@svsu.edu}\\[0.8in]
\large Department of Physics,\\ 
\large Saginaw Valley State University,\\ 
\large 7400 Bay Road University Center,  MI 48710 USA\\

\end{center}

\vspace{.5in}

\begin{abstract}
We consider the open spin-$s$ XXZ quantum spin chain with nondiagonal boundary terms. By exploiting certain functional relations at roots of unity, 
we derive a generalized form of $T$-$Q$ relation involving more than one independent $Q(u)$, which we use to propose the Bethe-ansatz-type 
expressions for the eigenvalues of the transfer matrix. At most two of the boundary parameters are set to be arbitrary and the bulk 
anisotropy parameter has values $\eta = {i \pi\over 2}\,, {i \pi\over 4}\,,\ldots $. We also provide numerical evidence for the completeness of the Bethe-ansatz-type 
solutions derived, using $s = 1$ case as an example.
\end{abstract}
\end{titlepage}

\setcounter{footnote}{0}

\section{Introduction}\label{sec:intro}

Considerable effort has been put into solving integrable quantum spin chains for many years. In particular, integrable open quantum spin chains
have attracted much interest over the years. In this regard, open XXX and XXZ quantum 
spin chains have been extensively investigated due to their growing applications in fields of physics such as statistical mechanics, string theory and 
condensed matter physics. Much progress has been made on the topic up to this point. Numerous successes in the 
past \cite{Gaudin}-\cite{Ma2} (also refer to \cite{Do1}-\cite{CRAb} and references therein, for other 
related work on the subject) have motivated further investigations of these models.  
In addition, in a series of publication, Bethe 
ansatz solutions have been derived for open spin-$1/2$ XXZ quantum spin chain where the boundary parameters obey a certain constraint. Readers are urged to refer 
to \cite{CLSW}-\cite{YZ1} for related work on the subject. Two sets of Bethe ansatz equations are needed there to obtain all $2^{N}$ 
eigenvalues, where $N$ is the number of sites. A special case of the above solution was generalized to open XXZ quantum spin chain with alternating spins by 
Doikou \cite{Do2} using the functional relation approach proposed by Nepomechie in \cite{Ne} to solve the spin-$1/2$ case. 
In \cite{Do3}, related work was carried out using the method in \cite{CLSW}. In \cite{FNR}, the spin-$1/2$ XXZ Bethe ansatz 
solution (for boundary parameters obeying certain constraint) is generalized to the spin-$s$ case by utilizing an approach based on the $Q$-operator and the $T$-$Q$ equation \cite{Baxter} (see below), 
which was developed earlier for the spin-$1/2$ XXZ chain in \cite{YNZ} and subsequently applied to the spin-$1/2$ XYZ chain in \cite{YZ3}. Two 
sets of Bethe ansatz equations are also needed there to produce all $(2s+1)^{N}$ eigenvalues, where again $N$ represents the number of sites. This was later followed by 
another work for spin-$s$ with such constraint removed, but limiting the Bethe ansatz solutions for cases with at most two arbitrary boundary parameters for some special 
values of the bulk anisotropy parameter \cite{Murganspins}, namely $\eta = \frac{i\pi}{p+1}$, with $p$ being even integers \cite{Murganspins}.

In a number of works cited above, the well known Baxter $T-Q$ relation \cite{Baxter}, with the following schematic form
\be
t(u)Q(u) = Q(v) + Q(w) 
\ee
has provided a way to obtain the Bethe ansatz equations for the eigenvalues of the transfer matrix $t(u)$. In \cite{MNS2, MNS2b}, a generalization of this relation
involving two $Q(u)$ of the following form for the open spin-$1/2$ XXZ quantum spin chain was given:
\be
t(u)Q_{1}(u) = Q_{2}(v) + Q_{2}(w)\non\\
t(u)Q_{2}(u) = Q_{1}(v') + Q_{1}(w')
\ee

Motivated by this solution, in this paper, we obtain the corresponding 
solution for the open spin-$s$ XXZ quantum spin chain. In addition, our work is motivated by the fact that such $T-Q$ relations for an open spin-$s$ XXZ quantum spin chain are 
novel structures and therefore merit further studies and investigation. We remark that a more general form of $T-Q$ relations were found in \cite{MNS3} involving
multiple $Q(u)$ functions. Moreover, the relation of $s = 1$ case to the supersymmetric sine-Gordon (SSG) model \cite{SSG}-\cite{SSGe}, especially the boundary SSG model 
\cite{BSSG}-\cite{MurganSSG}, has inspired us to consider the problem.  We stress that these results hold for cases with at most 
two arbitrary boundary parameters at roots of unity, namely when the bulk anisotropy parameter has vales $\eta = {i \pi\over 2}\,, {i\pi\over 4}\,,\ldots$. We follow 
similar approach as given in \cite{Ne}-\cite{NR} and \cite{MNS2} that was used to solve the $s = 1/2$ case, which is based on functional relations obeyed by transfer 
matrix at roots of unity. This yields Bethe-ansatz-type solutions which give the eigenvalues. Our numerical analysis for $s = 1$ case for $N = 2$ and $p = 3\,, 5$   
yields all the eigenvalues as given in Tables 1 and 2. As in \cite{FNR}, we rely on fusion \cite{MNR}, \cite{fusion}-\cite{fusion2b}, the truncation of the 
fusion hierarchy at roots of unity \cite{truncation}-\cite{truncationc} and the Bazhanov-Reshetikin \cite{BR} solution of the RSOS models.  

The outline of the paper is as follows: In Sec. 2, we review the construction of the fused $R$ \cite{fusion}-\cite{fusione}, \cite{Ka}-\cite{spinsXXZb} and $K^{\mp}$ \cite{MNR}, \cite{fusion2, fusion2b} 
matrices from the corresponding spin-$1/2$ matrices. One can refer to \cite{dVGR, GZ} for some original work on spin-$1/2$ $K^{\mp}$ matrices. We then review the 
construction of commuting transfer matrices from these fused matrices (using Sklyanin's work \cite{Sk}, which relies on Cherednik's previous results \cite{Ch}), 
together with some of their properties. Fusion hierachy and functional relations obeyed by transfer matrices are also reviewed. In Sec. 3, the generalized $T-Q$ relations are given 
along with some arguments behind their structure. This is done by exploiting the reviewed functional relations obeyed by the 
transfer matrices. From this, we derive the Bethe-ansatz-type equations for cases with at most 
two arbitrary boundary parameters at roots of unity, e.g. $\eta = {i \pi\over 2}\,, {i\pi\over 4}\,,\ldots$. We then present numerical results in Sec. 4 to illustrate the completeness of our 
solution, using $s = 1$ as an example. Here, the Bethe roots and 
energy eigenvalues derived from the Bethe-ansatz-type equations (for some values of $p$ and $N$) are given. We remark that these eigenvalues coincide with the ones obtained 
from direct diagonalization of the open spin-$1$ XXZ chain Hamiltonian. Finally, we conclude the paper with discussion of the results and potential future works in Sec. 5.

\section{Commuting spin-$s$ transfer matrices and functional relations at roots of unity}\label{sec:transfer}

In this section, in order to make the paper relatively self contained, we review some crucial concepts on the construction of commuting transfer matrices for $N$-site open spin-$s$ XXZ quantum spin chain. 
Materials reviewed here on fused $R$, $K^{\mp}$ and higher spin transfer matrices are mainly reproduced from \cite{FNR}. 
Like the commuting transfer matrix for $s=1/2$, constructed in \cite{Sk}, which we denote (following notations adopted in \cite{FNR}) by
$t^{(\frac{1}{2},\frac{1}{2})}(u)$, whose auxiliary space as well as each of
its $N$ quantum spaces are two-dimensional, one can construct a transfer matrix $t^{(j,s)}(u)$ whose
auxiliary space is spin-$j$ ($(2j+1)$-dimensional) and each of its $N$
quantum spaces are spin-$s$ ($(2s+1)$-dimensional), for any $j,s \in 
\{\frac{1}{2},1,\frac{3}{2},\ldots\}$ using the fused $R$ \cite{fusion}-\cite{fusione}, \cite{Ka}-\cite{spinsXXZb} and $K^{\mp}$ \cite{MNR}, \cite{fusion2, fusion2b} matrices. 
These matrices serve as building blocks in the construction of the commuting transfer matrices for higher spins. We list them 
below with some of their properties. The fused-$R$ matrices can be constructed as given below, 
\be
R^{(j,s)}_{\{a\} \{b\}}(u) =
P_{\{a\}}^{+} P_{\{b\}}^{+} 
\prod_{k=1}^{2j}\prod_{l=1}^{2s}
R^{(\frac{1}{2},\frac{1}{2})}_{a_{k} b_{l}}(u+(k+l-j-s-1)\eta)\, 
P_{\{a\}}^{+} P_{\{b\}}^{+} \,,
\label{fusedRmatrix}
\ee 
where $\{a\} = \{a_{1}, \ldots , a_{2j}\}$, $\{b\} = \{b_{1}, \ldots , 
b_{2s}\}$, and $P_{\{a\}}^{+}$ is the symmetric projector given by 
\be
P_{\{a\}}^{+} ={1\over (2j)!} 
\prod_{k=1}^{2j}\left(\sum_{l=1}^{k}{\cal P}_{a_{l}, a_{k}} \right) \,,
\label{projector}
\ee
${\cal P}$ is the permutation operator, with ${\cal P}_{a_{k}, 
a_{k}} \equiv 1$; similar definition also holds for $P_{\{b\}}^{+}$. 
$R^{(\frac{1}{2},\frac{1}{2})}(u)$ is given by
\be
R^{(\frac{1}{2},\frac{1}{2})}(u) = \left( \begin{array}{cccc}
	\sh  (u + \eta) &0            &0           &0            \\
	0                 &\sh  u     &\sh \eta  &0            \\
	0                 &\sh \eta   &\sh  u    &0            \\
	0                 &0            &0           &\sh  (u + \eta)
\end{array} \right) \,,
\label{Rmatrix}
\ee 
where $\eta$ is the bulk anisotropy parameter. 
The fused $R$ matrices satisfy the Yang-Baxter equations \cite{Yang1, Yang1b}
\be
R^{(j,k)}_{\{a\} \{b\}}(u-v)\, R^{(j,s)}_{\{a\} \{c\}}(u)\, 
R^{(k,s)}_{\{b\} \{c\}}(v) =
R^{(k,s)}_{\{b\} \{c\}}(v)\,  R^{(j,s)}_{\{a\} \{c\}}(u)\, 
R^{(j,k)}_{\{a\} \{b\}}(u-v) \,.
\ee 
The construction of the fused $K^{-}$ matrices now readily follows \cite{MNR}, \cite{fusion2, fusion2b},
\be
K^{- (j)}_{\{a\}}(u) &=& P_{\{a\}}^{+} \prod_{k=1}^{2j} \Bigg\{ \left[ 
\prod_{l=1}^{k-1} R^{(\frac{1}{2},\frac{1}{2})}_{a_{l}a_{k}}
(2u+(k+l-2j-1)\eta) \right] \non \\
&\times & K^{- (\frac{1}{2})}_{a_{k}}(u+(k-j-\frac{1}{2})\eta) \Bigg\}
P_{\{a\}}^{+} \,,
\label{fusedKmatrix}
\ee 
where $K^{- (\frac{1}{2})}(u)$ is the
$2 \times 2$ matrix whose components
are given by \cite{dVGR, GZ}
\be
K_{11}^{-}(u) &=& 2 \left( \sh \alpha_{-} \ch \beta_{-} \ch u +
\ch \alpha_{-} \sh \beta_{-} \sh u \right) \non \\
K_{22}^{-}(u) &=& 2 \left( \sh \alpha_{-} \ch \beta_{-} \ch u -
\ch \alpha_{-} \sh \beta_{-} \sh u \right) \non \\
K_{12}^{-}(u) &=& e^{\theta_{-}} \sh  2u \,, \qquad 
K_{21}^{-}(u) = e^{-\theta_{-}} \sh  2u \,,
\label{Kminuscomponents}
\ee
where $\alpha_{-} \,, \beta_{-} \,, \theta_{-}$ are the boundary
parameters.
The fused $K^{-}$ matrices satisfy the boundary Yang-Baxter equations 
\cite{Ch}
\be
\lefteqn{R^{(j,s)}_{\{a\} \{b\}}(u-v)\, K^{- (j)}_{\{a\}}(u)\,
R^{(j,s)}_{\{a\} \{b\}}(u+v)\, K^{- (j)}_{\{b\}}(v)}\non \\
& & =K^{- (j)}_{\{b\}}(v)\, R^{(j,s)}_{\{a\} \{b\}}(u+v)\,
K^{- (j)}_{\{a\}}(u)\, R^{(j,s)}_{\{a\} \{b\}}(u-v) \,.
\ee
In addition, the fused $K^{+}$ matrices are given by
\be
K^{+ (j)}_{\{a\}}(u)  = {1\over f^{(j)}(u)}\,K^{- (j)}_{\{a\}}
(-u-\eta)\Big\vert_{(\alpha_-,\beta_-,\theta_-)\rightarrow
(-\alpha_+,-\beta_+,\theta_+)} \,,
\ee
with the following normalization factor, 
\be
f^{(j)}(u) = \prod_{l=1}^{2j-1}\prod_{k=1}^{l}
[-\xi( 2u + (l+k+1-2j)\eta) ]\,. 
\label{Kplusnormalization}
\ee
From the fused matrices, one constructs the higher spin transfer matrix $t^{(j,s)}(u)$,
\be
t^{(j,s)}(u) = \tr_{\{a\}} K^{+ (j)}_{\{a\}}(u)\,
T^{(j,s)}_{\{a\}}(u)\, K^{- (j)}_{\{a\}}(u)\,
\hat T^{(j,s)}_{\{a\}}(u) \,.
\ee 
The monodromy matrices are given by products of $N$ fused $R$ 
matrices, 
\be
T^{(j,s)}_{\{a\}}(u) &=& R^{(j,s)}_{\{a\}, \{b^{[N]}\}}(u) \ldots 
R^{(j,s)}_{\{a\}, \{b^{[1]}\}}(u) \,, \non \\
\hat T^{(j,s)}_{\{a\}}(u) &=& R^{(j,s)}_{\{a\}, \{b^{[1]}\}}(u) \ldots
R^{(j,s)}_{\{a\}, \{b^{[N]}\}}(u) \,.
\ee 
These transfer matrices commute for different
values of spectral parameter for any $j , j' \in \{\frac{1}{2}, 1,
\frac{3}{2}, \ldots \}$ and any $s \in \{\frac{1}{2}, 1, \frac{3}{2},
\ldots \}$,
\be
\left[ t^{(j,s)}(u) \,, t^{(j',s)}(u') \right] = 0 \,.
\label{commutativity}
\ee
Furthermore, they also obey the fusion hierarchy \cite{MNR, FNR, fusion2, fusion2b}\footnote{See the appendix in \cite{FNR} for more details on the fusion hierachy.}  
\be
t^{(j-\frac{1}{2},s)}(u- j\eta)\, t^{(\frac{1}{2},s)}(u) =
t^{(j,s)}(u-(j-\frac{1}{2})\eta)  + \delta^{(s)}(u-\eta)\,
t^{(j-1,s)}(u-(j+\frac{1}{2})\eta) \,, 
\label{hierarchy}
\ee
$j = 1,\frac{3}{2},\ldots$, where $t^{(0,s)}=1$, and
$\delta^{(s)}(u)$ is given by
\be
\delta^{(s)}(u) &=& 
\delta_{0}^{(s)}(u)\delta_{1}^{(s)}(u)\,,
\label{dd}
\ee
where 
\be
\delta_{0}^{(s)}(u) &=& \left[\prod_{k=0}^{2s-1}\xi(u+(s-k+\frac{1}{2})\eta)\right]^{2N} 
{\sh(2u) \sh(2u+4\eta)\over \sh(2u+\eta) \sh(2u+3\eta)}\non \\
\delta_{1}^{(s)}(u) &=& 2^{4}\sh(u+\alpha_{-}+\eta)\sh(u-\alpha_{-}+\eta)\ch(u+\beta_{-}+\eta)\ch(u-\beta_{-}+\eta)\non \\
&\times& \sh(u+\alpha_{+}+\eta)\sh(u-\alpha_{+}+\eta)\ch(u+\beta_{+}+\eta)\ch(u-\beta_{+}+\eta) \,.
\label{delta01}
\ee
To avoid confusion, we emphasize that the $\delta^{(s)}(u)$ in \cite{FNR} differs from the one given here by a shift in $\eta$.

Next, we list a few important
properties of the rescaled ``fundamental'' transfer matrix
$\tilde t^{(\frac{1}{2},s)}(u)$ (defined below), which are useful in determining
its eigenvalues. Following the definition of $\tilde t^{(\frac{1}{2},s)}(u)$ as in \cite{FNR}, we have
\be
\tilde t^{(\frac{1}{2},s)}(u) = {1\over g^{(\frac{1}{2},s)}(u)^{2N}} 
t^{(\frac{1}{2},s)}(u) \,,
\label{tildet}
\ee
where
\be
g^{(\frac{1}{2},s)}(u) = \prod_{k=1}^{2s-1} \sh(u+(s-k+\frac{1}{2})\eta)\,.
\label{gfunction}
\ee
This transfer matrix has the following useful properties: 
\be
\tilde t^{(\frac{1}{2},s)}(u + i\pi) = \tilde 
t^{(\frac{1}{2},s)}(u) \qquad (i\pi\mbox{ - periodicity}) 
\label{periodicity}
\ee
\be 
\tilde t^{(\frac{1}{2},s)}(-u -\eta) = \tilde 
t^{(\frac{1}{2},s)}(u) \qquad (\mbox{crossing}) 
\label{crossing}
\ee
\be
\tilde t^{(\frac{1}{2},s)}(0) = -2^{3}\sh^{2N}((s+\frac{1}{2})\eta) 
\ch \eta \sh \alpha_{-} \ch \beta_{-} \sh \alpha_{+} \ch \beta_{+} \id
\quad (\mbox{initial condition} )
\label{initial}
\ee
\be
\tilde t^{(\frac{1}{2},s)}(u)\Big\vert_{\eta=0} &=& 
2^{3}\sh^{2N}u \Big[ -\sh \alpha_{-} \ch \beta_{-} \sh \alpha_{+} \ch 
\beta_{+} \ch^{2}u  \non \\
&+& \ch \alpha_{-} \sh \beta_{-} \ch \alpha_{+} \sh 
\beta_{+} \sh^{2}u \non \\
&-& \ch(\theta_{-}-\theta_{+}) \sh^{2}u \ch^{2}u 
\Big] \id \quad (\mbox{semi-classical limit} )
\label{semiclassical}
\ee
where $\id$ is the identity matrix.

Due to the commutativity property (\ref{commutativity}), the corresponding simultaneous eigenvectors are independent of the spectral 
parameter. Hence, (\ref{periodicity}) - (\ref{semiclassical}) hold for the corresponding eigenvalues as well. In addition to the above 
mentioned properties, for bulk anisotropy parameter values $\eta = {i \pi\over p+1}$, with $p= 1 \,, 2 \,, \ldots $, the ``fundamental'' transfer matrix, $t^{(\frac{1}{2},s)}(u)$ 
(and hence each of the corresponding eigenvalues, $\Lambda^{(\frac{1}{2},s)}(u)$) obeys functional relations of order $p+1$ \cite{Ne}-\cite{Nec},
\be
\lefteqn{t^{(\frac{1}{2},s)}(u) t^{(\frac{1}{2},s)}(u +\eta) \ldots t^{(\frac{1}{2},s)}(u + p \eta)} \non \\
&-& \delta^{(s)} (u-\eta) t^{(\frac{1}{2},s)}(u +\eta) t^{(\frac{1}{2},s)}(u +2\eta) 
\ldots t^{(\frac{1}{2},s)}(u + (p-1)\eta) \non \\
&-& \delta^{(s)} (u) t^{(\frac{1}{2},s)}(u +2\eta) t^{(\frac{1}{2},s)}(u +3\eta)
\ldots t^{(\frac{1}{2},s)}(u + p \eta) \non \\
&-& \delta^{(s)} (u+\eta) t^{(\frac{1}{2},s)}(u) t^{(\frac{1}{2},s)}(u +3\eta) t^{(\frac{1}{2},s)}(u +4\eta) 
\ldots t^{(\frac{1}{2},s)}(u + p \eta) \non \\
&-& \delta^{(s)} (u+2\eta) t^{(\frac{1}{2},s)}(u) t^{(\frac{1}{2},s)}(u +\eta) t^{(\frac{1}{2},s)}(u +4\eta) 
\ldots t^{(\frac{1}{2},s)}(u + p \eta) - \ldots \non \\
&-& \delta^{(s)} (u+(p-1)\eta) t^{(\frac{1}{2},s)}(u) t^{(\frac{1}{2},s)}(u +\eta) 
\ldots t^{(\frac{1}{2},s)}(u +  (p-2)\eta) \non \\
&+& \ldots  = f(u) \,.
\label{funcrltn}
\ee 
The scalar function $f(u)$ (which can be expressed as $f(u) = f_{0}(u)f_{1}(u)$) is given in terms of the boundary parameters $\alpha_{\mp} \,, \beta_{\mp} \,, \theta_{\mp}$ 
(for odd $p$) by  
\be
f_{0}(u)  = \left\{ 
\begin{array}{ll}
    (-1)^{N+1} 2^{-4 s p N}\sh^{4sN} \left( (p+1)u \right)\tnh^{2} \left( (p+1)u \right)\,, \\
    \qquad \qquad s= {1\over 2}\,, {3\over 2}\,, {5\over 2}\,, \ldots \\
   (-1)^{N+1} 2^{-4 s p N} \ch^{4sN} \left( (p+1)u \right)\tnh^{2} \left( (p+1)u \right)\,, \\
\qquad \qquad s= 1\,, 2\,, 3\,, \ldots \\
\end{array} \right.
\label{f0}
\ee
and
\be
f_{1}(u) &=& -2^{3-2 p} \Big( \non \\
& & \hspace{-0.2in}
\ch \left( (p+1) \alpha_{-} \right)\ch \left( (p+1) \beta_{-} \right)
\ch \left( (p+1) \alpha_{+} \right)\ch \left( (p+1) \beta_{+} \right)
\sh^{2} \left( (p+1)u \right) \non \\
&-&
\sh \left( (p+1) \alpha_{-} \right)\sh \left( (p+1) \beta_{-} \right)
\sh \left( (p+1) \alpha_{+} \right)\sh \left( (p+1) \beta_{+} \right)
\ch^{2} \left( (p+1)u \right) \non \\
&+&
(-1)^{N} \ch \left( (p+1)(\theta_{-}-\theta_{+}) \right)
\sh^{2} \left( (p+1)u \right) \ch^{2} \left( (p+1)u \right) 
\Big) \,,
\label{f1}
\ee
for $s = {1\over 2}\,, {3\over 2}\,, {5\over 2}\ldots$ 
and 
\be
f_{1}(u) &=& (-1)^{N+1} 2^{3-2 p} \Big( \non \\
& & \hspace{-0.2in}
\ch \left( (p+1) \alpha_{-} \right)\ch \left( (p+1) \beta_{-} \right)
\ch \left( (p+1) \alpha_{+} \right)\ch \left( (p+1) \beta_{+} \right)
\sh^{2} \left( (p+1)u \right) \non \\
&-&
\sh \left( (p+1) \alpha_{-} \right)\sh \left( (p+1) \beta_{-} \right)
\sh \left( (p+1) \alpha_{+} \right)\sh \left( (p+1) \beta_{+} \right)
\ch^{2} \left( (p+1)u \right) \non \\
&+&
 \ch \left( (p+1)(\theta_{-}-\theta_{+}) \right)
\sh^{2} \left( (p+1)u \right) \ch^{2} \left( (p+1)u \right) 
\Big) \,,
\label{f1ints}
\ee
for $s = 1\,, 2\,, 3\ldots$.
Note that $f(u)$ satisfies
\be
f(u + \eta) = f(u) \,, \qquad f(-u)=f(u) \,,
\ee
and
\be
f_{0}(u)^{2} = \prod_{j=0}^{p}\delta^{(s)}_{0}(u + j \eta) \,,
\label{identity}
\ee
where $\delta^{(s)}_{0}(u)$ is given by (\ref{delta01}).

\section{Generalized $T$-$Q$ relations and Bethe ansatz}

In this section, we give the main results of this paper. We derive the generalized $T-Q$ relations for the transfer matrix eigenvalues and obtain the Bethe-ansatz-type equations, 
for cases where at most two of the boundary parameters $\alpha_{\pm}$ or 
$\beta_{\pm}$ are arbitrary, by adopting the steps given in \cite{MNS2} while setting $\theta_{-} = \theta_{+}= \theta$, where $\theta$ is also arbitrary. 
More on this is given below.

\subsection{$T-Q$ relations}\label{subsec:TQ}

The transfer matrix $t^{(\frac{1}{2},s)}(u)$ and its eigenvalues ($\Lambda^{(\frac{1}{2},s)}(u)$) obey the functional relations (\ref{funcrltn}). We exploit this fact to 
obtain the $T-Q$ relations. Following \cite{BR}, one could recast the functional relations as the condition that the determinant  of a certain matrix vanishes, namely
\be
\det {\cal M}(u) = 0 \,,
\label{det}
\ee
where ${\cal M}(u)$ is given by the $(p+1) \times (p+1)$ matrix
\be
{\cal M} (u)= 
\left(
\begin{array}{cccccccc}
    \Lambda^{(\frac{1}{2},s)}(u) & -{\delta^{(s)}(u)\over h^{(1)}(u)} & 0  & \ldots  & 0 &
    -{\delta^{(s)}(u-\eta)\over h^{(2)}(u-\eta)}  \\
    -h^{(1)}(u) & \Lambda^{(\frac{1}{2},s)}(u+\eta) & -h^{(2)}(u+\eta)  & \ldots  & 0 & 0  \\
    \vdots  & \vdots & \vdots & \ddots 
    & \vdots  & \vdots    \\
   -h^{(2)}(u-\eta)  & 0 & 0 & \ldots  & -h^{(1)}(u+(p-1)\eta) &
    \Lambda^{(\frac{1}{2},s)}(u+p\eta) 
\end{array} \right)\non\\\,,
\label{newM}
\ee
where $h^{(1)}(u)$ and $h^{(2)}(u)$ are functions which are
$i\pi$-periodic, but otherwise not yet specified.
We note that the above matrix has the following symmetry,
\be
T \M(u) T^{-1} = {\cal M}(u+2\eta) \,, \qquad T \equiv S^{2} \,,
\label{symmetryT}
\ee
where $S$ is given by,
\be
S = \left(
\begin{array}{cccccccc}
    0 & 1 & 0  & \ldots  & 0 & 0  \\
    0 & 0 & 1  & \ldots  & 0 & 0  \\
    \vdots  & \vdots & \vdots & \ddots 
    & \vdots  & \vdots    \\
    0 & 0 & 0  & \ldots  & 0 & 1 \\
   1  & 0 & 0 & \ldots  & 0 & 0
\end{array} \right)\,.
\label{Tmatrix}
\ee
Assuming that 
\be
\det \M(u) =0 \,,
\label{vanishingdet}
\ee 
then $\M(u)$ has a null eigenvector $v(u)$,
\be
\M(u)\ v(u) = 0 \,.
\label{newnulleigenvector}
\ee 
The symmetry (\ref{symmetryT}) is consistent with
\be
T\ v(u) = v(u+2\eta) \,,
\ee
which implies that $v(u)$ has the form 
\be
v(u) = \left( Q_{1}(u)\,, Q_{2}(u+\eta) \,, \ldots \,, Q_{1}(u-2\eta) 
\,, Q_{2}(u-\eta)\right) \,, 
\label{newv} 
\ee
with
\be 
Q_{1}(u) = Q_{1}(u+i\pi) \,, \qquad  Q_{2}(u) = Q_{2}(u+i\pi) \,.
\label{Qperiodicity}
\ee
That is, the components of $v(u)$ are determined by {\it two} 
independent functions, $Q_{1}(u)$ and $Q_{2}(u)$.\footnote{In \cite{Murganspins}, all of the matrices $\M(u)$ possess a stronger symmetry, $S \M(u) S^{-1} = {\cal M}(u+\eta)$, 
implying the null eigenvector with single $Q(u)$. We refer the reader to Sec. 3 of \cite{MNS2} for more detail discussion on this.}  

The null eigenvector condition (\ref{newnulleigenvector})
together with the explicit forms of $\M(u)$ and $v(u)$, given by (\ref{newM}) and (\ref{newv}) respectively,
now lead to the following $T-Q$ relations,
\be
\Lambda^{(\frac{1}{2},s)}(u) &=& 
{\delta^{(s)}(u)\over h^{(1)}(u)} {Q_{2}(u+\eta)\over Q_{1}(u)} 
+ {\delta^{(s)}(u-\eta)\over h^{(2)}(u-\eta)} {Q_{2}(u-\eta)\over 
Q_{1}(u)} \,, \label{TQ1} \\
 &=& 
h^{(1)}(u-\eta) {Q_{1}(u-\eta)\over Q_{2}(u)} 
+ h^{(2)}(u) {Q_{1}(u+\eta)\over 
Q_{2}(u)} \,.
\label{TQ2}
\ee
Due to the crossing symmetry (\ref{crossing}) and 
\be
\delta^{(s)}(u) =  \delta^{(s)}(-u-2\eta)\,,
\label{deltaproperties}
\ee
which is the crossing property for $\delta^{(s)}(u)$, it is then natural to have 
the two terms in (\ref{TQ1}) transform into each other under crossing.
Hence, we set
\be
h^{(2)}(u) = h^{(1)}(-u-2\eta) \,,
\label{h2}
\ee
and we make the following ansatz
\be
Q_{j}(u) &=& \prod_{k=1}^{M_{j}} 
\sinh (u - u_{k}^{(j)}) \sinh (u + u_{k}^{(j)} + \eta) \,, 
\label{ansatz}
\ee 
which is consistent with the required periodicity (\ref{Qperiodicity})
and crossing properties
\be
Q_{j}(u) &=& Q_{j}(-u-\eta) \,,
\ee
where $j = 1\,, 2$. In (\ref{ansatz}), $\{u_{k}^{(j)}\}$ represents the Bethe roots (or zeros of $Q_{j}(u)$) and there are $M_{j}$ of these roots. 
Further, one can verify that the condition $\det \M(u)=0$ indeed implies the
functional relations (\ref{funcrltn}), if $w(u)$ satisfies
\be
f(u) = w(u) \prod_{j=0,2,\ldots}^{p-1} \delta^{(s)}(u+j\eta) 
+ {1\over w(u)} \prod_{j=1,3,\ldots}^{p} \delta^{(s)}(u+j\eta) \,,
\label{newconditiononh}
\ee
where 
\be
w(u) \equiv {\prod_{j=1,3,\ldots}^{p} h^{(2)}(u+j\eta)\over 
\prod_{j=0,2,\ldots}^{p-1} h^{(1)}(u+j\eta)} \,.
\label{wdefinition}
\ee
It follows from (\ref{newconditiononh}) that the process of finding $w(u)$ reduces to solving a quadratic equation, which when used together with 
(\ref{h2}) and (\ref{wdefinition}), yields the explicit form for the function  $h^{(1)}(u)$. Here, we consider even number of sites, $N$. Below, we give
the solutions of (\ref{wdefinition}) for $h^{(1)}(u)$ for two cases:

I. $\beta_{-}$ and $\beta_{+}$ arbitrary while setting $\alpha_{\pm} = 0$, $\theta_{-} = \theta_{+} = \theta$ = arbitrary.

\be
h^{(1)}(u) = 4\left[\prod_{k=0}^{2s-1}\sh(u+(s-k+\frac{3}{2})\eta)\right]^{2N} 
{\sh^2(u+\eta) \sh(2u+4\eta)\over \sh(2u+3\eta)} \,, \non\\ 
M_{1} = s N + \frac{1}{2}(p + 1) \,, \quad M_{2} = M_{1} - 1  \,.
\label{hbeta}
\ee

II. $\alpha_{-}$ and $\alpha_{+}$ arbitrary while setting $\beta_{\pm} = 0$, $\theta_{-} = \theta_{+} = \theta$ = arbitrary.

\be
h^{(1)}(u) = 4\left[\prod_{k=0}^{2s-1}\sh(u+(s-k+\frac{3}{2})\eta)\right]^{2N} 
{\ch^2(u+\eta) \sh(2u+4\eta)\over \sh(2u+3\eta)}\,, \non\\
M_{1} = s N + \frac{1}{2}(p + 1) \,, \quad M_{2} = M_{1} - 1  \,.
\label{halpha}
\ee

Now, using the analyticity of $\Lambda^{(\frac{1}{2},s)}(u)$, given by (\ref{TQ1}) and (\ref{TQ2}), one can write down the 
Bethe-ansatz-type equations for the zeros $\{ u_{j}^{(1)} \,,
u_{j}^{(2)} \}$ of $Q_{1}(u)$ and $Q_{2}(u)$,
\be
{\delta^{(s)}(u_{j}^{(1)})\ h^{(2)}(u_{j}^{(1)}-\eta)
\over \delta^{(s)}(u_{j}^{(1)}-\eta)\ h^{(1)}(u_{j}^{(1)})} 
&=&-{Q_{2}(u_{j}^{(1)}-\eta)\over Q_{2}(u_{j}^{(1)}+\eta)} \,, \qquad j =
1\,, 2\,, \ldots \,, M_{1} \,, \label{BAEa} \\
{h^{(1)}(u_{j}^{(2)}-\eta)\over h^{(2)}(u_{j}^{(2)})}
&=&-{Q_{1}(u_{j}^{(2)}+\eta)\over Q_{1}(u_{j}^{(2)}-\eta)} \,, \qquad j =
1\,, 2\,, \ldots \,, M_{2} \,.
\label{BAE}
\ee 
We remark here that for each case, there are more than one solutions for $h^{(1)}(u)$ that correspond to the above expression 
for $w(u)$. The solutions found are largely by trial and error, verifying numerically 
for small values of $N$ that the eigenvalues can indeed be expressed as 
(\ref{TQ1}), (\ref{TQ2}) with $Q(u)$'s of the form (\ref{ansatz}). 

To summarize, we have proposed that for the case where the bulk anisotropy parameter, $\eta = \frac{i\pi}{p+1}$ with $p$ being odd integers and 
that at most two of the boundary parameters are arbitrary,
the eigenvalues $\Lambda^{(\frac{1}{2},s)}(u)$ of the transfer matrix $t^{(\frac{1}{2},s)}(u)$ for two cases (I and II) are given by a generalized form 
of $T-Q$ relations (\ref{TQ1}), (\ref{TQ2}), with 
$Q_{1}(u)$ and $Q_{2}(u)$ given by (\ref{ansatz}) and $h^{(2)}(u)$ given 
by  (\ref{h2}). The $h^{(1)}(u)$ is given 
by (\ref{hbeta}) and (\ref{halpha}) respectively, for the two cases considered. The zeros $\{ u_{j}^{(1)} \,, u_{j}^{(2)} \}$ of
$Q_{1}(u)$ and $Q_{2}(u)$ are indeed the solutions of the Bethe-ansatz-type equations, (\ref{BAEa}) and (\ref{BAE}). These equations reproduce results in \cite{MNS2, MNS2b}
for $s=\frac{1}{2}$. In the following section, we shall use these results 
(specifically $\tilde{\Lambda}^{(\frac{1}{2},s)}(u)$ which represents the eigenvalues of the rescaled ``fundamental" transfer matrix given by (\ref{tildet})) to derive
expressions for energy eigenvalues for the case $s=1$.

\section{Energy eigenvalues and Bethe roots}

In this section, we provide numerical evidence for the completeness of the Bethe-ansatz-type solutions derived in Sec. 3 using the case $s=1$ as an example. 
We derive an expression for the energy eigenvalues for the open spin-$1$ XXZ quantum spin chain
and compute the complete energy levels for this case from the Bethe roots given by (\ref{BAEa}) and (\ref{BAE}). We do this for both cases, I and II. 

\subsection{Open spin-$1$ XXZ quantum spin chain: The Hamiltonian}

In this section, we review the integrable Hamiltonian for the open spin-1 XXZ quantum spin chain (adopting notations used in \cite{FNR}). 
The Hamiltonian is given by
\be
{\cal H} = \sum_{n=1}^{N-1}H_{n,n+1} + H_{b} \,,
\label{Hamiltonianspin1}
\ee
where $H_{n,n+1}$ represents the bulk terms. Explicitly, these terms are given by \cite{ZF},
\be 
H_{n,n+1} &=&  \sigma_{n} - (\sigma_{n})^{2}
+ 2 \sh^2 \eta \left[ \sigma_{n}^{z} + (S^z_n)^2
+ (S^z_{n+1})^2 - (\sigma_{n}^{z})^2 \right] \non \\
&-& 4 \sh^2 (\frac{\eta}{2})  \left( \sigma_{n}^{\bot} \sigma_{n}^{z}
+ \sigma_{n}^{z} \sigma_{n}^{\bot} \right) \,, \label{bulkhamiltonianspin1}
\ee 
where
\be
\sigma_{n} = \vec S_n \cdot \vec S_{n+1} \,, \quad
\sigma_{n}^{\bot} = S^x_n S^x_{n+1} + S^y_n S^y_{n+1}  \,, \quad
\sigma_{n}^{z} = S^z_n S^z_{n+1} \,, 
\ee 
and $\vec S$ are the $su(2)$ spin-1 generators. $H_{b}$ represents the boundary terms with the following form (see e.g., \cite{FNR, IOZ})
\be 
H_{b} &=& a_{1} (S^{z}_{1})^{2}  + a_{2} S^{z}_{1} 
+  a_{3} (S^{+}_{1})^{2}  +  a_{4} (S^{-}_{1})^{2}  +
a_{5} S^{+}_{1}\, S^{z}_{1}  + a_{6}  S^{z}_{1}\, S^{-}_{1} \non \\
&+& a_{7}  S^{z}_{1}\, S^{+}_{1} + a_{8} S^{-}_{1}\, S^{z}_{1} 
+ (a_{j} \leftrightarrow b_{j} \mbox{ and } 1 \leftrightarrow N) \,,
\ee
where $S^{\pm} = S^{x} \pm i S^{y}$. The coefficients $\{ a_{i} \}$ 
of the boundary terms at site 1 are functions  
of the boundary parameters ($\alpha_{-}, \beta_{-},
\theta_{-}$) and the bulk anisotropy parameter $\eta$. They are given by,
\be
a_{1} &=& \frac{1}{4} a_{0} \left(\ch 2\alpha_{-} - \ch 
2\beta_{-}+\ch \eta \right) \sh 2\eta 
\sh \eta \,,\non \\
a_{2} &=& \frac{1}{4} a_{0} \sh 2\alpha_{-} \sh 2\beta_{-} \sh 2\eta \,, \non \\
a_{3} &=& -\frac{1}{8} a_{0} e^{2\theta_{-}} \sh 2\eta 
\sh \eta \,, \non \\
a_{4} &=& -\frac{1}{8} a_{0} e^{-2\theta_{-}} \sh 2\eta 
\sh \eta \,, \non \\
a_{5} &=&  a_{0} e^{\theta_{-}} \left(
\ch \beta_{-}\sh \alpha_{-} \ch {\eta\over 2} +
\ch \alpha_{-}\sh \beta_{-} \sh {\eta\over 2} \right)
\sh \eta \ch^{\frac{3}{2}}\eta \,, \non \\
a_{6} &=&  a_{0} e^{-\theta_{-}} \left(
\ch \beta_{-}\sh \alpha_{-} \ch {\eta\over 2} +
\ch \alpha_{-}\sh \beta_{-} \sh {\eta\over 2} \right)
\sh \eta \ch^{\frac{3}{2}}\eta \,, \non \\
a_{7} &=&  -a_{0} e^{\theta_{-}} \left(
\ch \beta_{-}\sh \alpha_{-} \ch {\eta\over 2} -
\ch \alpha_{-}\sh \beta_{-} \sh {\eta\over 2} \right)
\sh \eta \ch^{\frac{3}{2}}\eta \,, \non \\
a_{8} &=&  -a_{0} e^{-\theta_{-}} \left(
\ch \beta_{-}\sh \alpha_{-} \ch {\eta\over 2} -
\ch \alpha_{-}\sh \beta_{-} \sh {\eta\over 2} \right)
\sh \eta \ch^{\frac{3}{2}}\eta \,,
\ee
where 
\be
a_{0}= \left[
\sh(\alpha_{-}-{\eta\over 2})\sh(\alpha_{-}+{\eta\over 2})
\ch(\beta_{-}-{\eta\over 2})\ch(\beta_{-}+{\eta\over 2})\right]^{-1} 
\,.
\ee
Similarly, the coefficients $\{ b_{i} \}$ of the boundary terms at 
site $N$ which are functions of  
the boundary parameters ($\alpha_{+}, \beta_{+}, \theta_{+}$) and $\eta$, are given by the following correspondence,
\be
b_{i} = a_{i}\Big\vert_{\alpha_{-}\rightarrow \alpha_{+}, 
\beta_{-}\rightarrow -\beta_{+}, \theta_{-}\rightarrow \theta_{+}} \,.
\ee
The Hamiltonian ${\cal H}$ (\ref{Hamiltonianspin1}), according to \cite{Sk}, is related to the first derivative of the spin-$1$ transfer matrix, namely $t^{(1,1)}(u)$, which one 
constructs from $t^{(\frac {1}{2},1)}(u)$ by using the fusion hierarchy formula (\ref{hierarchy}),
\be
t^{(1,1)}(u) = t^{(\frac {1}{2},1)}(u-{\eta\over 2})t^{(\frac {1}{2},1)}(u+{\eta\over 2}) - \delta^{(1)}(u-{\eta\over 2})\,,
\label{fh}
\ee
where
$\delta^{(1)}(u)$ is given by (\ref{dd})-(\ref{delta01}) with $s = 1$. Following \cite{FNR}, we work with the rescaled transfer matrix given by
\be
\tilde t^{(1,1)\ gt}(u) = {\sh(2u) \sh(2u+2\eta)\over [\sh u 
\sh(u+\eta)]^{2N}}\, 
t^{(1,1)\ gt}(u) \,,
\label{rescaled}
\ee
where $t^{(1,1)\ gt}(u)$ is the transfer matrix constructed from ``gauge''-transformed $R^{(1,1)}(u)$ and $K^{\mp(1)}(u)$ 
matrices \footnote{Such a transformation results in a more symmetric form of these matrices. For a detailed discussion on this, refer to Sec. 4 of \cite{FNR}.}. 
We note here that the rescaled transfer matrix does not vanish at $u=0$.

The Hamiltonian ${\cal H}$ (\ref{Hamiltonianspin1}), can now be expressed in terms of the first derivative of $\tilde t^{(1,1)\ gt}(u)$,
\be
{\cal H} = c^{(1)}_{1} {d \over du} \tilde t^{(1,1)\ gt}(u) \Big\vert_{u=0} 
+ c^{(1)}_{2} \id \,,
\label{firstderivatives1}
\ee
where 
\be 
c^{(1)}_{1}&=&\ch \eta \Big\{ 16 [\sh 2\eta \sh \eta]^{2N} \sh 3\eta 
\sh(\alpha_{-}-{\eta\over 2})\sh(\alpha_{-}+{\eta\over 2})
\ch(\beta_{-}-{\eta\over 2})\ch(\beta_{-}+{\eta\over 2})\non \\
&\times& \sh(\alpha_{+}-{\eta\over 2})\sh(\alpha_{+}+{\eta\over 2})
\ch(\beta_{+}-{\eta\over 2})\ch(\beta_{+}+{\eta\over 2})\Big\}^{-1}
\label{c1sone}
\,
\ee
and
\be 
c^{(1)}_{2}&=& -{a_{0}\over 4}b\ch\eta - (N-1)(4+\ch 2\eta) + 2 N \ch^{2}\eta \non \\
&-& {\sh\eta\over 2d}\Big\{-2\ch 2\alpha_{+}\Big(\ch\eta (3+7\ch 2\eta +\ch 4\eta)+\ch 2\beta_{+}(4+5\ch 2\eta+2\ch 4\eta)\Big)\non \\
&+& 2\ch \eta\Big(\ch 2\beta_{+}(3+7\ch 2\eta +\ch 4\eta)+\ch\eta (5+3\ch 2\eta +3\ch 4\eta)\Big)\Big\}\non \\
&-& {\sh 2\eta\over 2d}\Big\{\ch 2\beta_{+}(2+4\ch \eta \ch 3\eta)+\ch \eta (5\ch 2\eta +\ch 4\eta)-2\ch 2\alpha_{+}\Big(1+\ch 2\eta \non \\
&+& \ch 2\beta_{+}(\ch \eta +2\ch 3\eta)+\ch 4\eta\Big)\Big\}
\label{c2spin1}\,.
\ee
\noindent In (\ref{c2spin1}), $b$ and $d$ are given by
\be
b = 2\big(-\ch 2\beta_{-}-\ch^{3}\eta + \ch 2\alpha_{-}(1+\ch 2\beta_{-}\ch\eta)\big)  
\ee
and
\be
d = -4\sh 3\eta \sh(\alpha_{+}+{\eta\over 2})\sh(\alpha_{+}-{\eta\over 2})
\ch(\beta_{+}+{\eta\over 2})\ch(\beta_{+}-{\eta\over 2})\,.
\ee

\subsection{Open spin-1 XXZ quantum spin chain: Energy eigenvalues}

Next, we proceed to the eigenvalues of the Hamiltonian (\ref{firstderivatives1}).  Note that (\ref{firstderivatives1}) implies the following result for the corresponding 
eigenvalues,
\be
E = c^{(1)}_{1} {d \over du} \tilde \Lambda^{(1,1)\ gt}(u) \Big\vert_{u=0} 
+ c^{(1)}_{2}\,,
\label{energyfirstderivatives1}
\ee
where $\tilde \Lambda^{(1,1)\ gt}(u)$ represents the transfer matrix eigenvalues which assume the following form after using (\ref{tildet}), (\ref{fh}) and (\ref{rescaled}), 
\be
\tilde \Lambda^{(1,1)\ gt}(u) &=& {\sh(2u) \sh(2u+2\eta)\over [\sh u 
\sh(u+\eta)]^{2N}}\Big\{[g^{(\frac{1}{2},s)}(u-\frac{\eta}{2})g^{(\frac{1}{2},s)}(u+\frac{\eta}{2})]^{2N}\tilde \Lambda^{(\frac {1}{2},1)}(u-{\eta\over 2})\tilde \Lambda^{(\frac {1}{2},1)}(u+{\eta\over 2}) \non\\
&-& \delta^{(1)}(u-{\eta\over 2})\Big\}\,,
\label{fh2}
\ee
where
$\delta^{(1)}(u)$ is given by (\ref{dd})-(\ref{delta01}) with $s=1$. Furthermore, we have also used the fact that $\Lambda^{(1,1)\ gt}(u) = \Lambda^{(1,1)}(u)$. Finally, 
from (\ref{TQ2}), (\ref{fh}) and (\ref{rescaled}), we obtain the energy in terms of Bethe roots $\Large\{u_{k}^{(j)}\Large\}$ for cases I and II. Below, we present the
analytic forms of the energy eigenvalues for these two cases:

Case I. $\beta_{-}$ and $\beta_{+}$ arbitrary while setting $\alpha_{\pm} = 0$, $\theta_{-} = \theta_{+} = \theta$ = arbitrary :
\be
E &=& \sh^{2}(2\eta)\sum_{k=1}^{M_{1}}\frac{1}{\sh (u_{k}^{(1)}+{3\eta\over 2})\sh (u_{k}^{(1)} - {\eta\over 2})} 
 +2\sh 2\eta [(N+1)\cth \eta - \cth \frac{\eta}{2}] \non \\
&+& c^{(1)}_{1}C'(0)+c^{(1)}_{2}
\label{energyspinone1}
\ee

Case II. $\alpha_{-}$ and $\alpha_{+}$ arbitrary while setting $\beta_{\pm} = 0$, $\theta_{-} = \theta_{+} = \theta$ = arbitrary :
\be
E &=& \sh^{2}(2\eta)\sum_{k=1}^{M_{1}}\frac{1}{\sh (u_{k}^{(1)}+{3\eta\over 2})\sh (u_{k}^{(1)} - {\eta\over 2})} 
 +2\sh 2\eta [(N+1)\cth \eta - \tnh \frac{\eta}{2}] \non \\
&+& c^{(1)}_{1}C'(0)+c^{(1)}_{2}
\label{energyspinone2}
\ee
where in (\ref{energyspinone1}) and (\ref{energyspinone2}), 
\be
C(u) &=& -\frac{\sh 2u\sh(2u+2\eta)}{[\sh u \sh(u+\eta)]^{2N}}\delta^{(1)}(u - \frac{\eta}{2})\,.
\ee
We recall that $M_{1} = N + \frac{1}{2}(p+1)$ and $M_{2} = M_{1} - 1$. Note that in (\ref{energyspinone1}) and (\ref{energyspinone2}), the contribution to $E$
comes only from $\{u_{k}^{(1)}\}$ and not from both $\{u_{k}^{(1)}\}$ and $\{u_{k}^{(2)}\}$ as one may initially expect. Perhaps, if one uses (\ref{TQ1}) instead of 
(\ref{TQ2}) in the derivation of $E$, an equivalent expression involving only $\{u_k^{(2)}\}$ or both $\{u_{k}^{(1)}\}$ and $\{u_{k}^{(2)}\}$ may result. This however
will not affect the numerical value of the energy eigenvalues tabulated in the next section.

\subsection{Numerical results}

In this section, we tabulate the energies computed using (\ref{energyspinone1}) and (\ref{energyspinone2}) for some values of $N\,, p$ (therefore $\eta$) and the boundary parameters $\Large\{\alpha_{\pm}\,, \beta_{\pm}\,, \theta_{\pm}\Large\}$  
with the Bethe roots, $\Large\{u_{k}^{(1)}\Large\}$, in Tables 1 and 2 for cases I and II respectively. These Bethe roots are obtained using McCoy's method \cite{McCoy, McCoyb}. 
These numerical results demonstrate the completeness of the Bethe-ansatz-type 
equations, (\ref{BAEa}) and (\ref{BAE}) for the case $s = 1$. We checked these solutions for chains of length up to $N = 4$ for $p = 3\,,5$ and $7$ with boundary parameters $\beta_{+} = 0.695\,, \beta_{-} = 0.774$. 
We remark that these solutions reproduce the $s=\frac{1}{2}$ case given in \cite{MNS2, MNS2b}. The $T-Q$ relations (\ref{TQ1}) and (\ref{TQ2}) are also
numerically verified for $s=\frac{3}{2}$ case for some selected values of $N\,, p$ and boundary parameters. We acknowledge that these analysis provide 
some numerical support for the completeness of the Bethe-ansatz-type equations derived, (\ref{BAEa}) and (\ref{BAE}), and not a complete rigorous proof. We have also verified that the 
energies given in Tables 1 and 2 coincide with those obtained from direct diagonalization of (\ref{Hamiltonianspin1}).

\section{Discussion}\label{sec:discuss}

By using a method that relies on certain functional relations that the ``fundamental'' transfer matrices, $t^{(\frac{1}{2},s)}$(u), obey at roots of unity and the 
truncation of fusion hierarchy, we set up a generalized form of the $T-Q$ relation, (\ref{TQ1}) and (\ref{TQ2}), for the open spin-$s$ XXZ quantum spin chain with nondiagonal boundary 
terms. From these relations, we have determined Bethe-ansatz-type solutions of the model, (\ref{BAEa}) and (\ref{BAE}).  
These solutions hold only for $\eta = {i \pi\over 2}\,, {i \pi\over 4}\,,\ldots $. The solutions found here hold for arbitrary values of boundary parameters 
(at most two). These solutions have been checked for chains of length up to $N = 4$ for $p = 3\,,5$ and $7$ with boundary parameters $\beta_{+} = 0.695\,, \beta_{-} = 0.774$. 
We verified that they indeed produce all the $(2s + 1)^N$ transfer matrix eigenvalues for $s = \frac{1}{2}\,,1$ and $\frac{3}{2}$. Moreover, 
we also presented numerical evidence for the completeness of the Bethe-ansatz-type solutions found (using $s = 1$ as examples) in Tables 1 and 2. 
The numerical support for the completeness of the solutions presented here (using $s = 1$ case as examples) together with the results presented for spin-$1/2$ 
case in \cite{MNS2, MNS2b} and the fusion hierarchy (\ref{hierarchy}) which is used in the construction of higher spin-$s$ transfer matrices could perhaps
possibly enable one to develop a more formal rigorous proof for the completeness of the solutions found here. It would be interesting to pursue this in the 
future. 

In addition, a number of problems remain that are worth investigating. Perhaps one could carry out a more thorough treatment and analysis of the functional 
equation to yield the exact form of the $Q_{i}(u)$ functions, thus avoiding the need for an ansatz such as (\ref{ansatz}). Another interesting problem
is to see the relation of $s=1$ case to the supersymmetric sine-Gordon (SSG) 
model, along the lines of \cite{ANS} and \cite{MurganSSG}, but now for spin-$1$ chain with nondiagonal boundary terms described by the generalized $T-Q$ relations 
instead of the conventional $T-Q$ relation. One could also try to generalize the solutions presented in \cite{MNS3} for the spin-$1/2$ case, 
where all six boundary parameters are completely arbitrary, to any spin $s$, and analyze the $s=1$ case for this general solution in relation to the SSG model. In this 
regard, one can study the continuum limit of their Nonlinear Integral Equations (NLIEs), thus investigating the infrared (IR) and ultraviolet (UV) limits of the NLIEs. One could 
also investigate the boundary bound states of SSG models corresponding to all these cases such as reported recently in \cite{SSGDbound}. We hope to be able to address some of these 
issues in future publications.

\section*{Acknowledgments}

We thank the referees for their constructive and crucial comments and suggestions which greatly helped us in revising and improving the paper.

\newpage

\begin{table}[htb] 
	    \centering
	    \begin{tabular}{|c|c|c|}\hline
	     $E$ &  Bethe roots, $\Large\{u_{k}^{(1)}\Large\}$\\
	      \hline
	      -5.6483 & 0.426847 + 2.19193 i, 0.719676 + 1.1781 i, 0.109151 i,\\& 0.426847 + 0.164266 i\\    
	     -4.67715 & 0.106242 + 2.28424 i, 0.379199 + 1.1781 i, 1.05101 + 1.1781 i,\\& 0.106242 + 0.071957 i\\
                       -2.75841 & 0.387014 + 2.748893 i, 1.277532 i, 0.932369 + 1.1781 i,\\& 0.0609966 i\\
                       -1.98286 & 0.185547 + 2.748893 i, 1.701637 i, 0.915819 + 1.1781 i,\\& 0.138044 i\\
                       -1.54571 & 0.171807 + 3.046499 i, 0.171807 + 2.451287 i, 1.566925 i,\\& 0.916569 + 1.1781 i\\	
                       -0.489791 & 0.781754 + 1.921787i, 1.599981 i, 0.0312436 i,\\& 0.781754 + 0.434407 i\\
                       -0.392189 & 3.109568 i, 0.779636 + 1.920991 i, 1.554992 i,\\& 0.779636 + 0.435203 i\\
                       0.572634 & 0.810472 i, 0.624212 + 1.1781 i, 0.010646 i,\\& 1.227343 + 1.1781 i\\
                       0.808501 & 3.130312 i, 0.791507 i, 0.618753 + 1.1781 i,\\& 1.221033 + 1.1781 i\\
\hline
		  \end{tabular}
		  \caption[xxx]{\parbox[t]{0.8\textwidth}{
		  The 9 energies and corresponding Bethe roots 
		   for  
		  $N=2\,, s=1\,, p=3\,, \eta = i\pi/4\,,
		  \alpha_{-}=0\,, \beta_{-}=0.767\,, 
		  \theta_{-}=0.573\,, \alpha_{+}=0\,, 
		  \beta_{+}=0.598\,, \theta_{+}=0.573 
		  $}
		  }
		 \label{table:energiesM}
\end{table}  

\newpage

\begin{table}[htb] 
	    \centering
	    \begin{tabular}{|c|c|c|}\hline
	     $E$ &  Bethe roots, $\Large\{u_{k}^{(1)}\Large\}$\\
	      \hline
	      -6.07709 & 0.0471453 + 3.1415 i, 0.0471453 + 2.61809 i, 1.74867 i, 0.74532 + 1.309 i,\\& 0.48742 i\\    
	     -4.65604 & 2.65564 i, 0.107433 + 2.35618 i, 0.321204 + 1.309 i, 0.557414 i,\\& 0.107433 + 0.261819 i\\
                       -4.3506 & 0.00657235 + 3.07819 i, 0.00657235 + 2.6814 i, 2.07693 i, 0.12098 + 1.93837 i,\\& 0.12098 + 0.679624 i\\
                       -2.55991 & 0.272597 + 3.13706 i, 0.272597 + 2.62253 i, 2.13098 i, 0.672718 + 1.309 i,\\& 0.862768 i\\
                       -1.63092 & 0.326829 + 2.87979 i, 0.308315 + 2.35663 i, 2.13093 i, 0.890835 i,\\& 0.308315 + 0.261367 i\\	
                       0.0925845 & 0.248529 + 2.87979 i, 1.76311 i, 0.373083 + 1.309 i, 1.20497 + 1.309 i,\\& 0.487 i\\
                       0.0971716 & 0.548694 + 2.59187 i, 2.13099 i, 0.518481 + 1.309 i, 0.856853 i,\\& 0.548694 + 0.0261235 i\\
                       1.6757 & 0.70468 + 2.87979 i, 0.338436 + 1.309 i, 0.854426 i, 1.08306 + 1.309 i,\\& 0.487 i\\
                       2.99332 & 1.7639 i, 0.273003 + 1.309 i, 0.720682 + 1.309 i, 0.487 i,\\& 1.54847 + 1.309 i\\
\hline
		  \end{tabular}
		  \caption[xxx]{\parbox[t]{0.8\textwidth}{
		  The 9 energies and corresponding Bethe roots 
		  for  
		  $N=2\,, s=1\,, p=5\,, \eta = i\pi/6\,,
		  \alpha_{-}=0.854 i\,, \beta_{-}=0\,, 
		  \theta_{-}=0.482\,, \alpha_{+}=0.487 i\,, 
		  \beta_{+}=0\,, \theta_{+}=0.482 
		  $}
		  }
		 \label{table:energiesM}
\end{table}

\end{document}